\documentclass[reprint,superscriptaddress,amsmath,amssymb,aps,pra]{revtex4-2}

\usepackage{amsfonts}
\usepackage{amsmath}
\usepackage{amsthm}
\usepackage{mathrsfs}
\usepackage{amscd}
\usepackage{amssymb}
\usepackage{subfigure}
\usepackage{amsxtra}
\usepackage{dcolumn}
\usepackage{bm}           
\usepackage{bbm}
\usepackage{graphicx}
\usepackage{epstopdf}

\usepackage[colorlinks]{hyperref}
\usepackage[authormarkup=none]{changes}
\definechangesauthor[name={Stefan}, color=purple]{n}
\definechangesauthor[name={Stella}, color=blue]{s}
\definechangesauthor[name={Julia}, color=orange]{j}

\newcommand{\la}{\langle}
\newcommand{\ra}{\rangle}

\newcommand{\cD}{\mathcal{D}}
\newcommand{\cC}{\mathcal{C}}

\newcommand{\cI}{\mathit{I}}

\newcommand{\da}{\dagger}
\newcommand{\eps}{\varepsilon}
\newcommand{\Op}[1]{\hat{#1}}

\newcommand{\oA}{\Op{A}}

\newcommand{\vx}{\boldsymbol{x}}
\newcommand{\vtheta}{\boldsymbol{\theta}}
\newcommand{\veps}{\boldsymbol{\varepsilon}}
\newcommand{\vvtheta}{\boldsymbol{\vartheta}}

\newcommand{\diff}{\mathrm{d}}

\DeclareMathOperator{\argmin}{arg\,min}
\DeclareMathOperator{\argmax}{arg\,max}

\begin{document}

\preprint{APS/123-QED}

\title{Uninformed Bayesian Quantum Thermometry}
\author{Julia Boeyens}
\affiliation{Naturwissenschaftlich-Technische Fakult{\"a}t, Universit{\"a}t Siegen, Siegen 57068, Germany}

\author{Stella Seah}
\affiliation{D\'{e}partement de Physique Appliqu\'{e}e,  Universit\'{e} de Gen\`{e}ve,  1211 Gen\`{e}ve,  Switzerland}

\author{Stefan Nimmrichter}
\affiliation{Naturwissenschaftlich-Technische Fakult{\"a}t, Universit{\"a}t Siegen, Siegen 57068, Germany}

\date{\today}
             
\begin{abstract}
We study the Bayesian approach to thermometry with no prior knowledge about the expected temperature scale, through the example of energy measurements on fully or partially thermalized qubit probes. We show that the most common Bayesian estimators, namely the mean and the median, lead to high-temperature divergences when used for uninformed thermometry. To circumvent this and achieve better overall accuracy, we propose two new estimators based on an optimization of relative deviations. Their global temperature-averaged behavior matches a modified van Trees bound, which complements the Cram\'er-Rao bound for smaller probe numbers and unrestricted temperature ranges. 
Furthermore, we show that, using partially thermalized probes, one can increase the range of temperatures to which the thermometer is sensitive at the cost of the local accuracy.  
\end{abstract}
\maketitle

\section{Introduction}
Quantum thermometry \cite{Pasquale2018,Mehboudi2019a} is the current prime example of open quantum metrology. It comprises methods to infer the temperature of a thermal reservoir by probing it with small quantum systems and exploiting the quantum nature of the probes and the measurement scheme to enhance the precision of the temperature estimate. The advantages of these methods have found experimental applications in weakly invasive in-vitro thermometry of microscopic organisms~\cite{Fujiwara2020} as well as low-temperature monitoring in ultracold quantum gases \cite{leanhardt2003cooling,Ruostekoski2009,bloch2012quantum,Mehboudi2019,Carcy2021}.
 
Most theoretical considerations have focused on equilibrium thermometry in the asymptotic limit of large measurement data so that the Cram\'{e}r-Rao bound applies \cite{DeP2016,DeP2017,Hovhannisyan2018,Razavian2019,Bouton2020,Gebbia2020,Zhang2020}, mostly relying on qubits as the paradigmatic quantum probes \cite{Brunelli2012,Jevtic2015b,Cavina2018,Mitchison2020}. Strategies for improved temperature precision, as quantified in terms of the thermal Fisher information \cite{Pasquale2018,Mehboudi2019a}, consist in tailoring the energy spectrum of high-dimensional probes for a desired temperature range \cite{correa2015individual}, optimizing the binning of measurement outcomes \cite{Potts2019,Jorgensen2020,Hovhannisyan2021}, or using a catalyst \cite{Henao2021}.

In practice however, thermal equilibrium between the reservoir and the probes may not always be achievable \cite{Guo2015,Correa2016,Sekatski2021}, or even desirable \cite{Stace2010,Hofer2017b,Pati2020,Henao2021a}. In fact, the transient dynamics induced by repeated finite-time collisions with nonequilibrium quantum probes can result in enhanced precision compared to equilibrium thermometry \cite{Kiilerich2018,Seah2019a,Feyles2019}, once again quantified by the Fisher information. 

While the Fisher information is the key figure of merit in most studies, its predictive power is limited to the ideal scenario when measurements can be repeated many times. In the opposite, less explored, regime when only scarce data is available, the Bayesian parameter estimation framework is more appropriate. It encodes any information or bias known beforehand into a \emph{prior} distribution of expected temperatures, which is then updated by the measured outcome into a \emph{posterior} distribution from which to infer the temperature estimate. This allows one to describe both single uninformed measurement runs as well as optimized adaptive protocols. Applied before in Heisenberg-limited phase estimation \cite{Higgins2007,Wiebe2016,Yan2018,Gebhart2021}, the Bayesian formalism was only recently introduced in the context of thermometry \cite{rubio2020global,alves2020bayesian,Jorgensen2021,mehboudi2021fundamental}. 

Here we study uninformed Bayesian estimation from a single measurement outcome, assuming a fixed measurement scheme, but no further knowledge about the expected reservoir temperature. Results are evaluated for sequences of identical qubits probing the reservoir. We elaborate on the crucial role of the chosen estimator and prior distribution for the attainable accuracy within the range of detectable temperatures, which can be inferred from the likelihood function. In particular, given that the moments of the posterior temperature distribution typically do not exist, common choices such as the mean estimator necessitate temperatures be restricted to a finite range, which implies that the thermometry protocol cannot be truly uninformed. We propose two new estimators based on relative deviations in the temperature that alleviate the divergence problem and generally give better estimates with lower uncertainty and bias. Moreover, we compare fully thermalized to partially thermalized probes, showing that the latter are sensitive to a wider range of temperatures at the expense of the maximum achievable accuracy.

This paper is organised as follows: Sec.~\ref{sec:bayes} details the Bayesian approach to thermometry, including the prior distribution and all the estimators and measures of uncertainty that we consider. We then focus on our case study of qubit probes in and out of thermal equilibrium in Sec.~\ref{sec:model}, followed by a numerical benchmark assessment of the various estimators in Sec.~\ref{sec:results}. Specifically, for fully thermalised probes, we compare the average estimates and errors as functions of the true temperature in Sec.~\ref{sec:equilResults}, the average performance as a function of the probe number in Sec.~\ref{sec:globalScaling}, and also the influence of the prior in Sec.~\ref{sec:priorCompare}. Finally, we demonstrate the advantage of partially thermalized qubits in Sec.~\ref{sec:noneqResults}, before we conclude in Sec.~\ref{sec:conclusion}. 

\section{Bayesian Framework}
\label{sec:bayes}

In the uninformed Bayesian approach to thermometry \cite{Lehmann1998,jaynes2003probability}, there is no prior knowledge about the temperature $T$ of the thermal reservoir in question, apart from its positivity and its influence on the outcomes of the chosen measurement scheme. We assume a single measurement that gives one of $N+1$ different outcomes, labeled by $n=0,\ldots N$. The achievable temperature sensitivity is encapsulated in the \emph{likelihood function}: the conditional probability $P(n|T)$ to observe outcome $n$ given the temperature value $T$, as predicted by the theoretical model of the measurement procedure. 

Typically, one can associate a characteristic energy scale $E$ to the measurement scheme and thus express temperature in units of $E/k_B$.
Temperature sensitivity could be optimized by adjusting $E$ if prior information about the expected temperature range were available \cite{correa2015individual}, but we seek to infer $T \geq 0$ without that knowledge at fixed $E$. Bayesian parameter estimation theory requires us to make additional assumptions about (i) the \emph{prior} $P^{(0)} (T)$ that describes the ``expected'' distribution of temperatures prior to any measurement, (ii) a \emph{cost function} $c(\theta_n , T) \geq 0$ quantifying how ``wrong'' the temperature estimate $\theta_n$ based on the observed outcome $n$ is when the true temperature value was $T$, and (iii) a suitable \emph{error measure} that quantifies the uncertainty around the temperature estimate $\theta_n$ without reference to the inaccessible true value $T$.

For the temperature estimator itself, $\vtheta = (\theta_0,\ldots \theta_N)$, there exists a natural, optimal choice determined by (i) and (ii): the argument $\vvtheta = \argmin_{\vtheta} \cC(\vtheta)$ that minimizes the average cost over all possible true temperatures $T$ and outcomes $n$,
\begin{eqnarray}
   \cC(\vtheta) &=&  \int_0^{\infty} \!\! \diff T\, P^{(0)}(T) \sum_{n=0}^N P(n|T) c(\theta_n, T) \nonumber \\
    &=& \sum_{n=0}^N P(n) \int_0^{\infty} \!\! \diff T\, P(T|n) c(\theta_n, T).  \label{eq:costFunc}
\end{eqnarray}
In the second line, we have applied Bayes' rule and expressed the average in terms of the normalized \emph{posterior} 
\begin{eqnarray} \label{eq:posterior}
    P(T|n) &=& \frac{P(n|T)P^{(0)}(T)}{P(n)}, \nonumber \\
    P(n) &=& \int_0^\infty \!\! \diff T \, P(n|T)P^{(0)}(T),
\end{eqnarray}
which encodes the updated expectation about the distribution of temperatures after the measurement outcome $n$ was obtained. Here we include only possible outcomes, assuming $P(n)>0$. Moreover, given the infinite range of possible $T$-values, one must choose an appropriate prior $P^{(0)}(T)$ so as to ensure that all outcome probabilities $P(n)$ be finite. Moments of temperature, $\la T^k\ra = \int\diff T\, P(T|n) T^k$, might however still diverge.

\subsection{The Jeffreys prior}

For the estimation of a single continuous parameter $T$, it is standard practice to assume Jeffreys' prior \cite{Lehmann1998,Ghosh2011}. It represents the least informative starting point, because it maximizes the information gain in terms of the relative entropy between the prior and the posterior and it is invariant under re-parametrizations $T \to f(T)$. Jeffreys' prior is given by the square root of the Fisher information of the likelihood function with respect to the parameter,
\begin{equation}
    P^{(0)}(T) \propto \sqrt{\cI(T)} = \sqrt{ \sum_n P(n|T) \left[\partial_T \ln P(n|T)\right]^2}.
\end{equation}
The Fisher information $I(T)$ expresses the local temperature resolution of the given measurement scheme. Accordingly, Jeffreys' prior distribution assigns more weight to those temperatures at which the scheme is most sensitive. Other distributions could be chosen for convenience, but they would lead to additional bias by restricting the temperature range or by favoring temperatures that cannot be detected that well. When data is limited, the choice of prior affects the temperature estimator and error measure noticeably, as we will exemplify in Sec.~\ref{sec:priorCompare}. 

Moreover, the norm of the prior might not always be finite, which poses no problem in practice if the relevant posteriors \eqref{eq:posterior} and estimates $\vartheta_n$ remain well-defined. However, this is not the case for simple examples like the constant flat prior, or the scale-invariant prior $P^{(0)}(T) \propto 1/T$, and one must therefore restrict them to a finite temperature interval in practice. 

\subsection{Cost functions and estimators}\label{sub:estimators}

The cost function $c(\theta,T)$ measures the penalty of wrong temperature estimates by assigning a positive number to any deviation from the true $T$-value. One may choose to penalize deviations differently depending on their relative or absolute size, which then determines the optimal estimator $\vvtheta$ as well as an associated error measure $\veps$ \cite{jaynes2003probability}. Note that certain combinations of prior and $c(\theta,T)$ can result in diverging average costs \eqref{eq:costFunc}, estimates $\vartheta_n$, or errors $\eps_n$, and should therefore be avoided. In our case study, we consider six different estimators: (md) the mode of the posterior, (2) the absolute mean, (2r) the relative mean, (1) the median, (1r) the relative median, and (2l) the logarithmic mean. 

A widely used choice is the maximum-likelihood estimator with estimates $\vartheta_n^{(\rm{ml})} = \argmax_\theta P(n|\theta) $. It follows by assuming the (improper) flat prior $P^{(0)}(T) \propto 1$ and minimizing the average over the singular cost function $c^{(\rm{ml})}(\theta_n,T) = -\delta(\theta_n - T)$, which penalizes \emph{any} finite deviation by the same amount. 
For the Jeffreys prior used here, or for any other prior that restricts the admitted temperature range, the estimator denoted by  $\vvtheta^{\rm{(md)}}$ is rather given by the mode of the posterior, $\vartheta_n^{(\rm{md})} = \argmax_\theta P(\theta|n) $.

Another common cost function penalizes square absolute deviations, $c^{(2)}(\theta_n,T) = (\theta_n - T)^2$, which makes the mean of the posterior the optimal estimator, denoted $\vvtheta^{(2)}$ with estimates $\vartheta_n^{(2)} = \la T\ra$. The error can be measured in multiples of the corresponding standard deviation, provided that the posterior has finite first and second moments, which will however not be the case here.

A viable alternative is obtained by defining the cost function in terms of the \emph{relative} square deviation, $c^{(\rm{2r})}(\theta_n,T) = (\theta_n/T - 1)^2$, which penalizes deviations relative to the absolute temperature scale. This results in the relative mean estimator $\vvtheta^{\rm{(2r)}}$, the estimates of which will not diverge due to unrestricted temperatures, $\vartheta_n^{\rm{(2r)}} = \la T^{-1}\ra/\la T^{-2}\ra$.

The problem of diverging moments is also partly alleviated by using the absolute median estimator $\vvtheta^{(1)}$, which merely assumes that the posteriors \eqref{eq:posterior} be normalizable. It optimizes the average over the 1st-moment cost function $c^{(1)}(\theta_n,T) = |\theta_n - T|$, which would still diverge for an unrestricted temperature range. Nevertheless, the $N+1$ estimates $\vartheta_n^{(1)}$ remain finite and are defined through the identity
\begin{equation} \label{eq:median}
    \int_0^{\vartheta_n^{(1)}} \!\! \diff T \, P(T|n) = \int_{\vartheta_n^{(1)}}^\infty \!\! \diff T \, P(T|n) = \frac{1}{2}.
\end{equation}
Notably, $\vvtheta^{(1)}$ is invariant under any re-parametrization $T \to f(T)$, including simple rescaling and typical conversion formulas from temperature to excitation numbers. The median estimator shares this feature with the maximum-likelihood estimator using a flat prior, and with the credibility-based error \eqref{eq:error90} below. 

Once again, we could decide to penalize relative deviations instead and divide the cost function by the absolute temperature,  $c^{\rm{(1r)}}(\theta_n,T) = |\theta_n/T - 1|$, circumventing the problem of diverging $T$-moments. This results in the relative median estimator $\vvtheta^{\rm{(1r)}}$ with estimates $\vartheta^{\rm{(1r)}}_n$ given by the median of the re-normalized posterior distributions $\propto P(T|n)/T$.

One obtains a different class of estimators if one converts $T$ into a derived, physically motivated parameter $f(T)$ and applies one of the above standard cost functions in this new parameter space. Natural parametrizations of temperature would be the fermionic or bosonic excitation number, or simply $f(T) = \ln T$. The latter was recently proposed in combination with the square deviation $c^{\rm{(2l)}} (\theta_n, T) = \ln^2 (\theta_n/T)$ in log-temperature space, which results in the logarithmic mean estimator $\vvtheta^{\rm{(2l)}}$ with estimates 
\begin{equation}\label{eq:RubioEstimator}
    \vartheta_n^{\rm{(2l)}} = \frac{E}{k_B} \exp \left[ \int \!\! \diff T\, P(T|n) \ln \left( \frac{k_B T}{E}\right)\right],
\end{equation}
given a reference energy scale $E$ \cite{rubio2020global}. The authors employed the $1/T$-prior (i.e.~flat prior in $\ln T$), which necessitated a high- and low-temperature cutoff to guarantee finite estimates; Jeffreys' prior alleviates this issue.

\subsection{Uncertainty and error measures}

In the asymptotic large-data limit $N\to \infty$, the Bernstein-von Mises theorem states that the posterior \eqref{eq:posterior} will be sharply peaked like a Gaussian around the true $T$-value and thus the temperature estimates based on the various mentioned estimators should all eventually converge to that true value \cite{Vaart1998,LeCam}. 

At small $N$ however, the posterior distribution remains broad and influenced by the prior, leading to a high degree of uncertainty as well as likely deviations from the true value. It is therefore crucial to associate errors $\eps_n$ to the estimated $\vartheta_n$, which faithfully reproduce the actual uncertainty about the true temperature without knowing it. In fact, the precise (and often biased) $\vartheta_n$-values are not relevant so long as the corresponding uncertainty around them is large. 

In Bayesian single-parameter estimation, uncertainty can be universally measured in terms of posterior quantiles: parameter values $\theta_n^{X\%}$ at which the cumulative posterior distribution reaches a certain percentage level,
\begin{equation}\label{eq:quantile}
    \int_0^{\theta_n^{X\%}} \!\!\diff T \, P(T|n) \stackrel{!}{=} X\%.
\end{equation}
Quantiles are parametrization-invariant if Jeffreys' prior is used, they are independent of the chosen estimator, and they exist whenever the posterior has finite norm. We can reasonably claim with, say, $90\%$ credibility that the true temperature lies within the $5\%$- and the $95\%$-quantile for measured $n$, $\theta_n^{5\%} \lesssim T \lesssim \theta_n^{95\%}$. This suggests a temperature uncertainty of
\begin{equation}\label{eq:error90}
    \eps_n^{90\%} (\vartheta_n) = \theta_n^{95\%} - \theta_n^{5\%} 
\end{equation}
regardless of the choice of estimator $\vvtheta$. The so defined credibility region reflects what temperatures the experimenter deems possible with 90\% probability after obtaining a single measurement outcome $n$.

Bayes' rule links the credibility region around $\vartheta_n$ to the corresponding confidence interval $[\vartheta_a,\vartheta_b]$ of estimated temperatures, as determined by the greatest integers $a,b$ such that the cumulative sum of likelihoods yields $\sum_{n=0}^a P(n|T) \leq 5\%$ and $\sum_{n=0}^a P(n|T) \leq 95\%$ at a given true $T$ \cite{Jaynes1976, jaynes2003probability}. The confidence interval depends on the chosen estimator and reflects the range of temperature estimates an experimenter would obtain in 90\% of measurement instances on a reservoir at temperature $T$.
A well-behaved prior and estimator would ensure that the 90\%-credibility and the 90\%-confidence intervals are comparable and cover both the estimate $\vartheta_n$ and the true temperature $T$ for any possible outcome $n$ and detectable $T$-value.

Our numerical case study in Sec.~\ref{sec:results} reveals that the credibility-based error \eqref{eq:error90} tends to overestimate the actual deviation between estimated and true temperature whenever the latter exceeds the reference scale $E/k_B$. Hence, \eqref{eq:error90} is a rather conservative error measure, and errors based on the estimators' underlying cost functions are a convenient and possibly more accurate alternative.

To this end, consider the posterior-averaged cost $\la c(\vartheta_n, T) \ra$ given outcome $n$.
It quantifies how much the estimated $\vartheta_n$ could still deviate on average from the unknown true temperature. If the cost function $c(\vartheta_n, T)$ is already in units of temperature, then an appropriate multiple of its posterior average (divided by $\vartheta_n$) can directly serve as an absolute (or relative) error measure. Otherwise, the average must be translated to an associate temperature scale first. For the cost functions considered here, we can define the temperature errors as 
\begin{eqnarray} \label{eq:errors}
    \eps_n^{(1)} &=& 4.12\, \la c^{(1)} (\vartheta_n^{(1)}, T) \ra, \\ 
    \eps_n^{\rm{(1r)}} &=& 4.12\, \vartheta_n^{\rm{(1r)}} \la c^{\rm{(1r)}} (\vartheta_n^{\rm{(1r)}}, T) \ra , \nonumber \\
    \eps_n^{(2)} &=& 3.29\, \sqrt{ \la c^{(2)} (\vartheta_n^{(2)}, T) \ra }, \nonumber \\ 
    \eps_n^{\rm{(2r)}} &=& 3.29\, \vartheta_n^{\rm{(2r)}} \sqrt{ \la c^{\rm{(2r)}} (\vartheta_n^{\rm{(2r)}}, T) \ra }, \nonumber \\
    \eps_n^{\rm{(2l)}} &=& 2 \vartheta_n^{\rm{(2l)}} \sinh \left[ \frac{3.29}{2} \sqrt{ \la c^{\rm{(2l)}} (\vartheta_n^{\rm{(2l)}}, T) \ra } \right] \nonumber
\end{eqnarray}
The singular cost function underlying the estimator $\vvtheta^{\rm{(md)}}$ does not produce a meaningful error measure. For fair comparison, the prefactors in the first four lines are chosen such that the errors all agree with the $90\%$-credibility measure \eqref{eq:error90} for Gaussian posteriors sharply peaked around the estimated temperatures, as expected in the large-$N$ limit. For the logarithmic estimator $\vvtheta^{\rm{(2l)}}$, the posterior-averaged cost function gives the variance $\sigma_n^2 = \la c^{\rm{(2l)}} (\vartheta_n^{\rm{(2l)}}, T) \ra$ around the estimated $\ln \vartheta_n^{\rm{(2l)}}$ in $\ln T$-space. We obtain from this a 90\%-matched temperature interval by converting the two boundary values $\ln \vartheta_n^{\rm{2l}} \pm 3.29 \sigma_n/2$ to $T$-space and taking the difference.

\section{Qubit thermometry}
\label{sec:model}

\begin{figure}
\centering
\includegraphics[width=1.0\columnwidth]{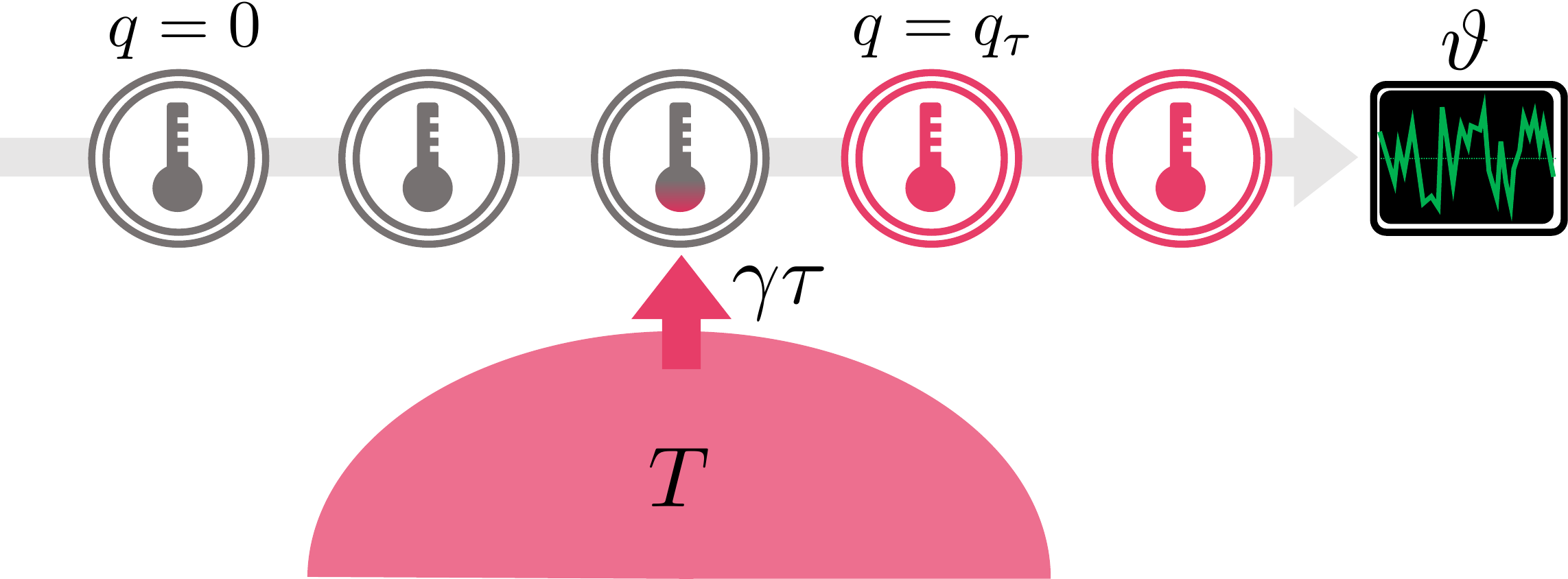}
\caption{\label{fig:sketch}Sketch of the qubit thermometer. $N$ ground-state qubits exchange heat with a bosonic reservoir at temperature $T$ and rate $\gamma$, each for a duration $\tau$. This partial thermalization brings the qubit ensemble to a mixed state with excitation probability $q_{\tau}$. The experimenter infers the temperature from the measured number of excitations by means of a Bayesian estimator $\vvtheta$.}
\end{figure}

For our case study, we consider the energy-based qubit thermometry setting sketched in Fig.~\ref{fig:sketch}, in which a bosonic thermal reservoir of temperature $T$ is successively probed by $N$ identical qubits with energy gap $E$. Each probe exchanges heat with the reservoir through a weak thermal contact over a fixed duration $\tau$, after which it is decoupled again and replaced with the next qubit probe. Once every probe has undergone its partial equilibration with the reservoir, the experimenter measures the number $n=0,\ldots N$ of excitations in the ensemble. 
We assume that the qubit probes are prepared in their ground state,  which corresponds to the most temperature-sensitive full-swap protocol of collisional thermometry studied in \cite{Seah2019a}. Correlations between subsequent probes as studied in \cite{Shu2020} are precluded here.

The thermal coupling is described by the standard master equation for a qubit state $\rho$ in an oscillator bath, with rate parameter $\gamma$. In the rotating frame, 
\begin{equation}
    \dot{\rho} = \frac{\gamma}{1-e^{-E/k_B T}} \left( \cD[|0\ra\la 1|]\rho + e^{-E/k_B T} \cD[|1\ra\la 0|]\rho \right),
\end{equation}
with $\cD[\oA]\rho = \oA\rho\oA^\da - \{\oA^\da\oA,\rho\}/2$, and $|0\ra$ and $|1\ra$ the qubit's ground and excited state. Crucially, the master equation predicts an enhanced effective thermalization rate of approximately $\gamma k_B T/E$ in the high-temperature regime $k_B T \gg E$. After the coupling time $\tau$, the initial ground state evolves into a mixture of ground and excited state with excitation probability 
\begin{equation}\label{eq:qtau}
q_\tau (T) = \frac{1 - e^{-\gamma \tau \coth (E/2k_B T)} }{1+e^{E/k_B T}}. 
\end{equation}
In the limit of long coupling times, $\gamma \tau \gg 1$, the probes equilibrate to the Gibbs state, $q_\infty (T) = 1/(1+e^{E/k_B T})$. Measuring the number of excitations in the probe ensemble then amounts to an equilibrium thermometry scheme with $N$ repetitions. 

Measurement sensitivity degrades exponentially in the low-temperature limit $k_B T \ll E$ where $q_\tau (T) \sim e^{-E/k_B T} $ converges to zero. At high temperatures and thermal equilibrium, the excitation probability saturates at $1/2$ and the sensitivity also degrades, $ q_\infty (T) \approx 1/2 - E/4k_B T$ for $k_B T \gg E$. In contrast, a finite time $\tau$ yields 
\begin{equation} \label{eq:qtau_highT}
    q_\tau (T) \xrightarrow{k_B T \gg E} \left( \frac{1}{2} - \frac{E}{4k_B T} \right) \left(1-e^{-2\gamma \tau k_B T/E} \right),
\end{equation}
which de-saturates excitations at sufficiently small $\gamma \tau $. Indeed, we will discuss the sensitivity range of the thermometer more thoroughly in Sec.~\ref{sec:sensitivity range} and demonstrate the resulting increase in high-temperature accuracy for small $\gamma\tau$ in Sec.~\ref{sec:noneqResults} below.

\subsection{Likelihood and Jeffreys' prior}

The excitation probability of each qubit probe represents an independent coin toss with ``winning'' probability $q_\tau (T)$, and the likelihood for $n$ excitations in $N$ trials follows a Bernoulli chain,
\begin{equation}
    P_\tau(n|T) = \binom{N}{n} q_\tau^n (T) \left[ 1 - q_\tau (T) \right]^{N-n}, \label{eq:Bernoulli} 
\end{equation}
which converges to a fair coin toss in the asymptotic limit $T\to \infty$. 
The corresponding Jeffreys prior does not depend on the number of trials,
\begin{equation} \label{eq:JeffreyQubit}
    P^{(0)}_\tau (T) = \frac{2 \partial_T q_\tau (T)}{\pi \sqrt{q_\tau (T) [1-q_\tau(T)]}}.
\end{equation}
We omit the lengthy expression of the $T$-derivative here. The prior is correctly normalized, but already its first moment in $T$ diverges, because of the asymptotic behaviour $P_\tau^{(0)} (T) \sim T^{-2}$ for $T \to \infty$. 
Crucially, this implies that the first posterior moments $\la T\ra$ as well as the higher ones diverge for any outcome $n$, such that the mean estimator $\vvtheta^{(2)}$ and the error measures $\veps^{(1)}$ and $\veps^{(2)}$ are no longer well-defined. If one tries to remove the divergence by restricting the allowed temperature range to $T \in [0,T_{\max}]$, then the respective estimates and errors will become sensitive to the chosen upper bound $T_{\max}$. Hence the mean estimator $\vvtheta^{(2)}$ and the two moment-based error measures $\veps^{(1)}$ and $\veps^{(2)}$ are not suitable for noninformative qubit thermometry. 
The median estimator $\vvtheta^{(1)}$ in combination with the credibility region $\veps^{90\%}$, on the other hand, would not diverge, and neither do the relative estimators and associated errors in Sec.~\ref{sub:estimators}.

Picking a simpler prior with worse asymptotic behaviour should be avoided as it adds to the complications. For example, the $1/T$-prior cannot be normalized, and it invalidates not only the mean estimator $\vvtheta^{(2)}$, but also the median estimate $\vartheta^{(1)}_0$ when zero excitations are measured, due to divergence at $T\to 0$. Alternatively, if a finite $T$-range is imposed a priori, the estimates will be sensitive to the chosen temperature bounds. 
Estimation of bounded parameters, as studied in quantum phase metrology, is typically not plagued by such problems.

\subsection{Sensitivity range}
\label{sec:sensitivity range}

While the uninformed approach does not exclude temperatures prior to measurement, any thermometer will only be accurate within a certain temperature range determined by the likelihood function. In the qubit case, temperatures much smaller (greater) than $E/k_B$ are no longer distinguishable from $T=0$ ($T=\infty$). 

We propose to estimate the sensible range of temperatures by comparing the likelihoods in the low- and high-temperature limit. Given the $N$ bits of information the measurement provides in our scenario, we deem two temperatures $T_1, T_2$ barely distinguishable if the relative base-2 entropy between their likelihoods, 
\begin{equation}
    D(T_1 \| T_2) = \sum_{n=0}^N P(n|T_1) \log_2 \frac{P(n|T_1)}{P(n|T_2)},
\end{equation}
measures no less than 1\,bit. It quantifies the amount of discriminating information for outcomes sampled from either likelihood. For the Bernoulli chain \eqref{eq:Bernoulli}, we get 
\begin{eqnarray}
    D(T_1 \| T_2) &=& N q_\tau (T_1) \log_2 \frac{q_\tau (T_1)}{q_\tau (T_2)}  \\
    &&+ N [1-q_\tau (T_1)] \log_2 \frac{1-q_\tau (T_1)}{1-q_\tau (T_2)}.\nonumber
\end{eqnarray}
It is zero only if $T_1=T_2$, and otherwise a positive number of at most $N$ bits. We can thus restrict our view on temperatures between the boundaries $T_0$ and $T_{\infty}$ at which $D(0 \| T_0) = 1$ and $D (\infty \| T_{\infty}) = 1$, that is, 
\begin{equation}\label{eq:Tbounds}
    q_\tau (T_0) = 1-2^{-1/N}, \quad q_\tau (T_{\infty}) = \frac{1-\sqrt{1-4^{-1/N}}}{2} .
\end{equation}
Temperatures outside this range will not be discernible by the $N$-probe measurement. 

\begin{figure}
\centering
\includegraphics[width=1.0\columnwidth]{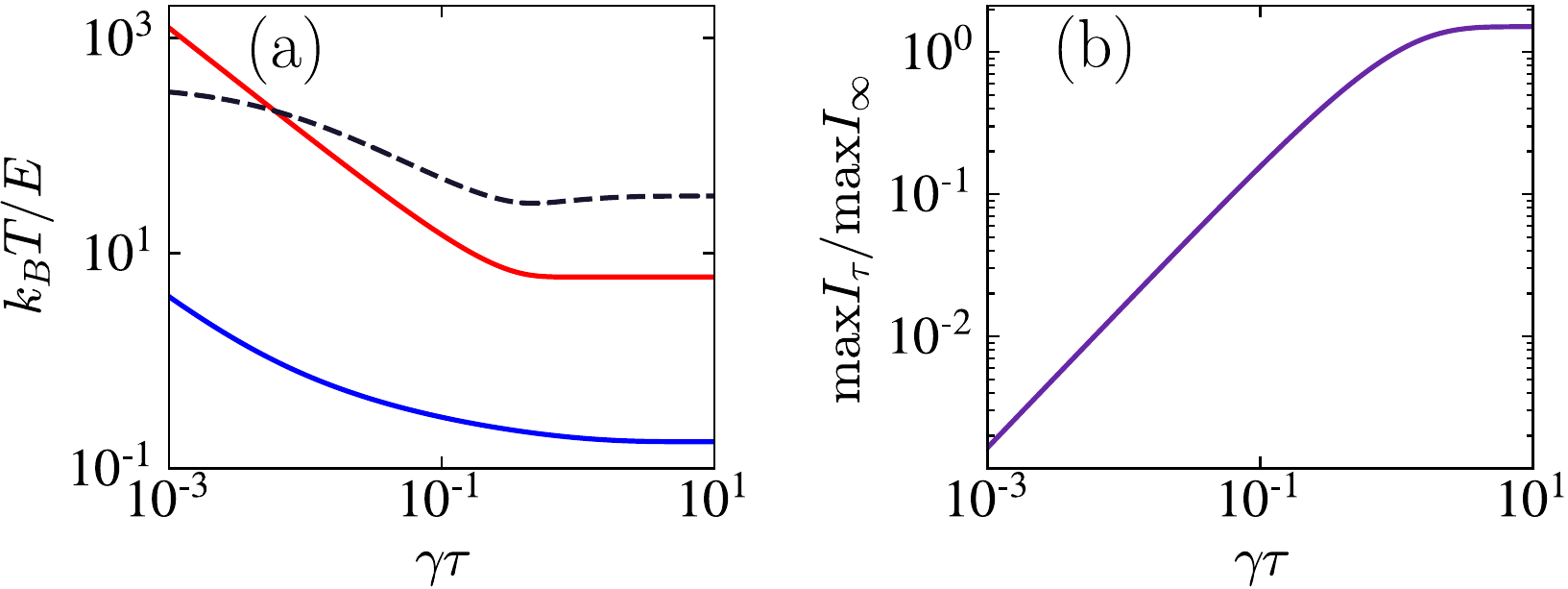}
\caption{\label{fig:Trange} (a) Lowest and highest detectable temperature, $T_0$ (blue, lower) and $T_{\infty}$ (red,upper) in units $E/k_B$, as well as $T_{\infty}/T_0$ (dashed) as a function of thermalization time $\gamma\tau$ for 200 qubits. (b) Highest attainable temperature resolution relative to equilibrium, $\max_T I_\tau (T)/\max_T I_\infty (T)$  versus $\gamma \tau$.}
\end{figure}

Figure \ref{fig:Trange}(a) depicts the boundary values $T_{\infty}$ and $T_0$ as well as the ratio between them as a function of $\gamma\tau$ for $N=200$ qubits. The right end of the diagram corresponds to the equilibrium thermometry limit, $\gamma \tau \gg 1$. We observe that non-equilibrium thermometry at $\tau < \infty$ shifts and extends the \emph{range} of detectable temperatures towards higher values with decreasing thermalization time $\tau$---a useful feature when nothing is known about the expected temperature scale a priori. 

However, the increased range comes at the price of reduced local temperature \emph{accuracy}, i.e.~lower sensitivity of the likelihood function to changes in the true $T$-value. This amounts to an overall lower Fisher information $I_\tau (T)$ of the likelihood with respect to $T$ for a given $\tau$. As a figure of merit, we plot the highest value relative to the equilibrium case, $\max_T I_\tau (T)/\max_T I_\infty (T)$ in Fig.~\ref{fig:Trange}(b). It decreases monotonically with decreasing $\gamma \tau$, which implies that the highest local temperature sensitivity is always achieved at equilibrium. 
Exemplary detailed comparisons between equilibrium and non-equilibrium estimation are provided in Sec.~\ref{sec:noneqResults}.

\begin{figure*}
\centering
\includegraphics[width=1.0\textwidth]{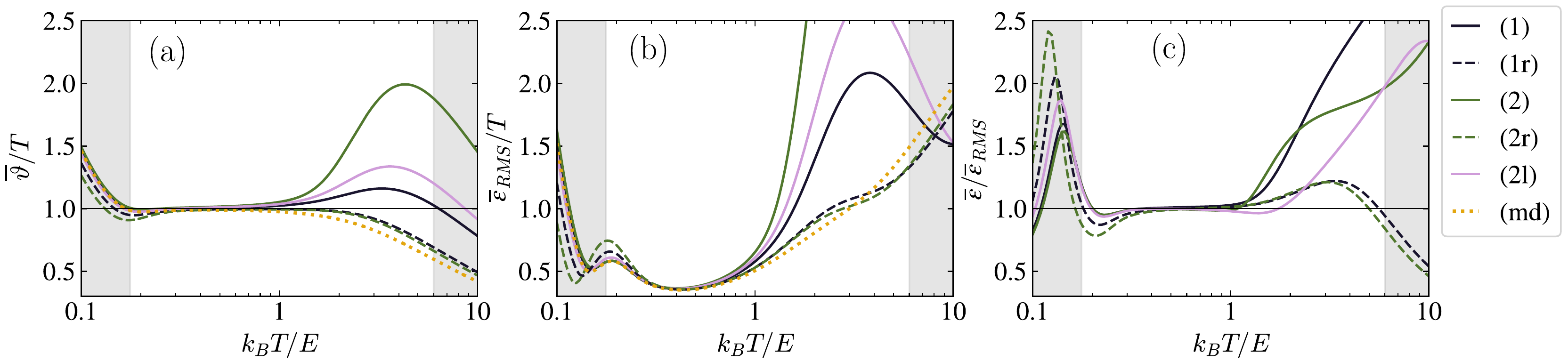}
\caption{\label{fig:comparison equlibrium} (Color online) Comparison of the Bayesian estimators from Sec.~\ref{sec:bayes} and their associated error measures \eqref{eq:errors} as a function of the true temperature $T$ for 200 fully thermalized qubits. 
(a) Outcome-averaged estimated temperatures $\bar\vartheta (T)$ from \eqref{eq:avgEstErr} relative to $T$; (b) average RMS deviations \eqref{eq:RMSerr} relative to $T$; (c) average error measures $\bar\eps (T)$ from \eqref{eq:avgEstErr} relative to the RMS deviations, excluding the mode estimator. The grey regions indicate temperatures outside the detectable range given by \eqref{eq:Tbounds}. }
\end{figure*}

Assuming tunable probe parameters, the trade-off between temperature range and local accuracy can be exploited in a two-stage measurement protocol: The first measurement stage would be carried out with a rapid sequence of qubits at small $\gamma \tau$ in order to efficiently narrow down the reservoir's temperature scale and choose an appropriate energy scale $E$ for optimal sensitivity in the second stage of equilibrium thermometry.

\section{Numerical assessment}\label{sec:results}

We assess the performance of different temperature estimators in qubit thermometry. For the most part, we consider thermal equilibrium, achieved in the limit of long qubit-reservoir coupling times, $\gamma\tau \gg 1$. Nonequilibrium results at shorter times are presented in Sec.~\ref{sec:noneqResults}.

Local and global figures of merit for the average performance over a broad range of temperatures are discussed in Secs.~\ref{sec:equilResults} and \ref{sec:globalScaling}, respectively.
Additionally, we assess the influence of different chosen priors in Sec.~\ref{sec:priorCompare}, showing that Jeffreys' prior yields more accurate estimates than biased priors. 

In our computations, we discretized the temperature support with a step size of $1 \cdot 10^{-3}$ in the range $k_B T/E \in [0.01,200]$, which was chosen so that the cutoffs were well outside of the sensitive range dictated by \eqref{eq:Tbounds}. 

\subsection{Comparison of estimators}\label{sec:equilResults}

We compare the various estimators introduced in Sec.~\ref{sec:bayes} in terms of the bias and the associated error measure. To this end, we shall employ the weighted averages over all outcomes at a given true temperature $T$, 
\begin{equation}\label{eq:avgEstErr}
    \bar\vartheta (T) = \sum_n P(n|T) \vartheta_n, \quad \bar\eps (T) = \sum_n P(n|T) \eps_n.
\end{equation}
In an actual  experiment, only one out of $N+1$ random outcomes would be observed, leading to a random estimate $\vartheta_n$. It is therefore crucial that the error measures $\eps_n$ accurately capture the spread of random outcomes, as quantified by the 90\% confidence interval, and the deviation from the true $T$-value. The latter can be measured in terms of the root-mean-square (RMS) deviation,
\begin{equation}\label{eq:RMSerr}
    \bar\eps_{\text{RMS}} (T) = 3.29\sqrt{\sum_n P(n|T) (\vartheta_n-T)^2},
\end{equation}
once again scaled to match 90\% credibility.

Figure \ref{fig:comparison equlibrium} plots (a) the outcome-averaged biases $\bar\vartheta (T)$, (b) the RMS deviations $ \bar\eps_{\text{RMS}} (T)$, and (c) the measured errors $\bar\eps (T)$ of the various estimators against $T$. The results were evaluated for a rather coarse measurement with $N=200$ qubits at equilibrium, and using Jeffreys' prior. The grey shaded regions mark temperatures outside the detectable range $[T_0,T_\infty]$ defined in Sec.~\ref{sec:sensitivity range}.

Considering Fig.~\ref{fig:comparison equlibrium}(a) alone, one might rule out the mean estimator $\vvtheta^{(2)}$ based purely on the large bias at high temperatures. However, it is unclear whether the median $\vvtheta^{(1)}$ and the logarithmic $\vvtheta^{\rm{(2l)}}$ are better than the relative estimators $\vvtheta^{\rm{(2r)}}$ and $\vvtheta^{\rm{(1r)}}$ or the mode $\vvtheta^{\rm{(md)}}$. 

Bias is not the most important figure of merit when assessing the estimators since it only provides information about how good the estimator is on average and not how much the estimator will vary from experiment to experiment. The latter is better represented by the average RMS deviation of the estimates relative to the true temperature, as shown in Fig.~\ref{fig:comparison equlibrium}(b). A similar performance is seen for all estimators in the intermediate temperature range $k_B T/E \lesssim 1$ at which the thermometer is most sensitive. At higher temperatures, on the other hand, the mean, the median, and the logarithmic mean perform significantly worse than the mode and the two relative estimators. The latter are slightly worse close to the low end of the detectable $T$-spectrum, but generally perform best overall. 

\begin{figure*}
\centering
\includegraphics[width=\textwidth]{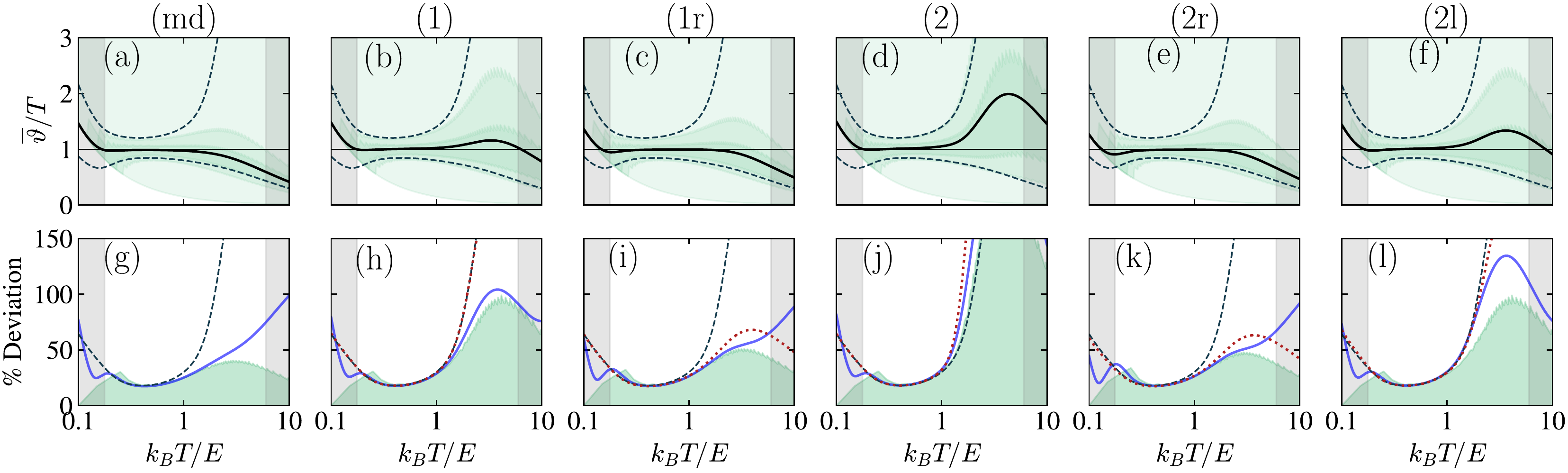}
\caption{\label{fig:gallery} Temperature estimates and errors for $N=200$ thermal qubits, complementing Fig.~\ref{fig:comparison equlibrium}. Each two-panel column shows the results for the respective Bayesian estimator referenced by the label on top: the mode in (a,g), the median in (b,h), the relative median in (c,i), the mean in (d,j), the relative mean in (e,k), and the logarithmic mean in (f,l). In the top row (a-f), the solid curves show the average biases $\bar \vartheta (T)$, the dashed lines delimit the 90\% credible regions, and the green shaded regions from light to dark mark the 100\%, 90\%, and 50\% confidence intervals, all relative to the true temperature $T$. 
The bottom row (g-l) shows the RMS deviation from $T$ (solid lines), the credible region widths $\bar\eps^{90 \%}$ (dashed), the average measured errors $\bar\eps(T)$ (dotted), and the 90\% confidence intervals (green shades), all given in \% with respect to $T$. The vertical grey bars delimit the detectable temperature range.}
\end{figure*}

A good estimator is of limited use if it lacks an associated error measure that the experimenter can evaluate without knowing the true $T$. The error measures discussed in \eqref{eq:errors} depend only on the posterior distribution and thus require no knowledge of the true temperature. For the estimator to be useful, the outcome-averaged associated error measure $\bar\eps (T)$ should match the actual deviation $\bar\eps_{\text{RMS}} (T)$ at a given true $T$. We plot the ratio of both quantities in Fig.~\ref{fig:comparison equlibrium}(c).
Once again, the relative estimators stand out over the whole detectable $T$-range. The mode estimator, which performs equally well in terms of bias and RMS deviation, lacks a meaningful error measure and is therefore absent. We will omit it for the rest of our analysis. 

Further detail is seen in Fig.~\ref{fig:gallery}. Here, the top row depicts the estimators' biases (solid curves), their 90\%-credible regions (between the dashed curves), as well as their associated 100\%, 90\%, and 50\% confidence intervals (light to dark green shades). In the bottom row, we compare the relative RMS deviations $\bar\eps_{\text{RMS}}(T)/T$ (solid) to the 90\% credibility ranges $\bar \eps^{90\%}$ (dashed), the measured errors $\bar\eps/T$ (dotted), and the 90\% confidence interval (green shade). 

Both the RMS deviations and the measured errors capture the actual spread of outcomes, i.e.~confidence intervals, quite well within the sensitive region of temperatures. The confidence intervals depend on the estimator and vary in size quite notably.
Unfortunately, the 90\%-credibility range $\bar \eps^{90\%}$ from \eqref{eq:error90}---a universal and robust error measure that does not depend on the chosen estimator---consistently overestimates the actual uncertainties at higher temperatures and is thus not a viable measure. 
The error measures associated with the median and the mean estimator also rise quickly at high $T$, and they would diverge if we included arbitrarily high temperatures in our assessment. Hence, their high-temperature values are unreliable as they vary with increasing numerical temperature cutoff. Considering all these limitations, we conclude that the relative mean and median are the best temperature estimators.

\subsection{Global error scaling}\label{sec:globalScaling}

Next, we discuss how the estimators' underlying temperature-averaged cost functions \eqref{eq:costFunc} and averaged errors scale with an increasing number of qubits. This has emerged as a figure of merit to quantify the ``global'' performance of temperature estimators in the Bayesian framework \cite{rubio2020global,alves2020bayesian}. Given that the $T$-moments diverge when integrating over all true $T\geq 0$, we shall restrict the integration to a finite range of relevant temperatures, $0< T_1 \leq T \leq T_2 < \infty$, for a fair comparison between all estimators. That is, we restrict \eqref{eq:costFunc} to
\begin{equation}\label{eq:finite_c}
    \cC_{\text{fin}} (\vtheta) =  \frac{1}{\mathcal{M}} \int_{T_1}^{T_2} \!\! \diff T\, P^{(0)}(T) \sum_{n=0}^N P(n|T) c(\theta_n, T),
\end{equation}
where we renormalize the prior \eqref{eq:JeffreyQubit} accordingly by 
\begin{equation}
    \mathcal{M} 
    = \frac{4}{\pi} \left[ \arcsin \sqrt{q_\tau (T_2)} - \arcsin \sqrt{q_\tau (T_1)} \right].
\end{equation}

\begin{figure}
\centering
\includegraphics[width=1.0\columnwidth]{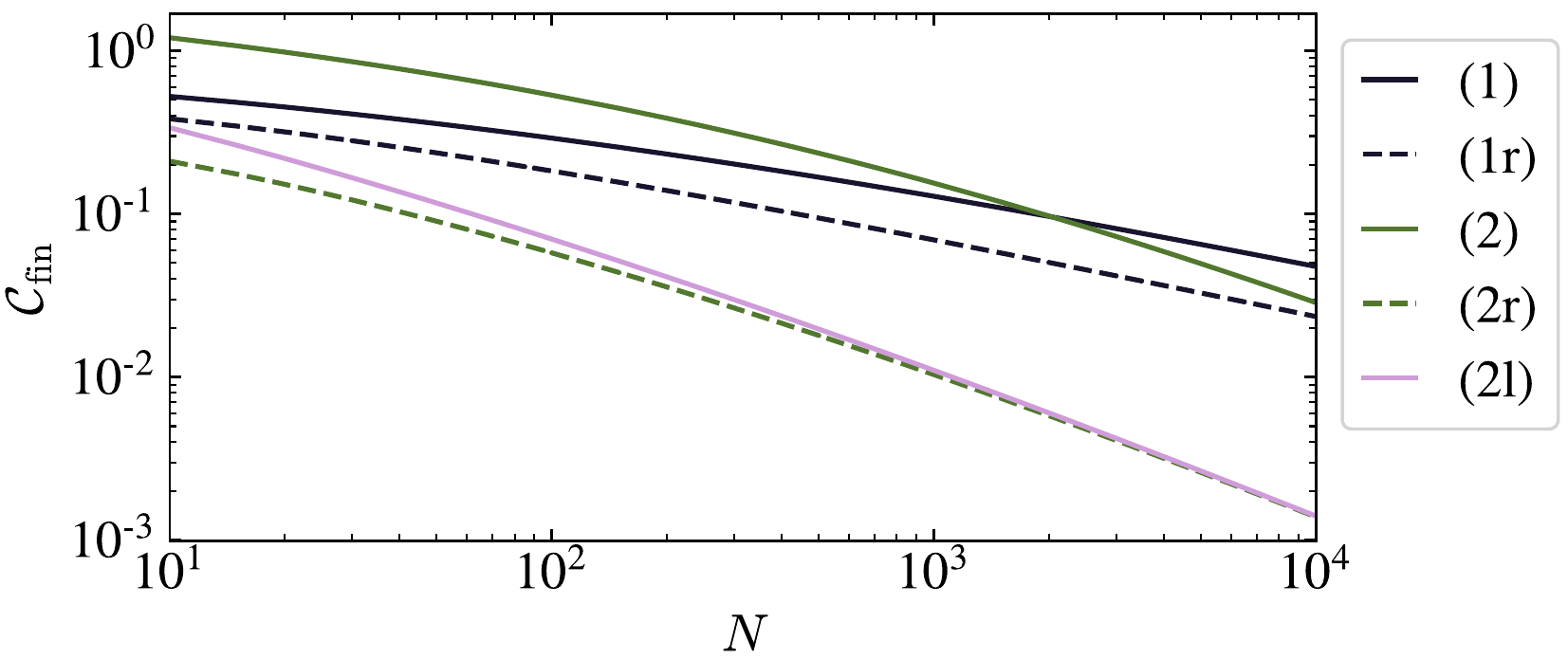}
\caption{\label{fig:scaling_cost} (Color online) Average costs \eqref{eq:finite_c} over the temperature range $k_B T/E\in [0.1,10]$ as a function of probe number $N$, evaluated for the listed estimators at thermal equilibrium. As the cost functions vary in dimension of $T$, we set $E/k_B \equiv 1$. }
\end{figure}

The scaling of our considered average cost functions with $N$ is shown in Fig.~\ref{fig:scaling_cost} for a temperature range covering two orders of magnitude around the reference scale, $k_BT/E\in [0.1,10]$, as depicted in the previous figures. Naively, one may think that, because the cost functions of the logarithmic and relative mean estimators scale best, their associated estimators perform better overall. However, this claim is simply invalid since the plotted cost functions are all based on different parameterizations in $T$ and thus differ in dimension. 

For a proper global comparison of the overall estimated uncertainty from low to high temperatures at given $N$, it is more reasonable to take the measured \emph{relative} temperature errors $\eps_n/\vartheta_n$ and average them over all outcomes, 
\begin{equation} \label{eq:globalE}
    \mathcal{E}_{\text{fin}}(\mathbf{\vvtheta}) =\sum_{n=0}^N P(n) \frac{\eps_n}{\vartheta_n}.
\end{equation}
Here, the sub-index in $\mathcal{E}_{\text{fin}}$ denotes that the individual outcomes $(\vartheta_n,\eps_n)$ taken from \eqref{eq:errors} be evaluated consistently, assuming the same restricted temperature range $[T_1,T_2]$ when computing the posterior \eqref{eq:posterior}. The resulting quantities $\mathcal{E}_{\text{fin}}$ are dimensionless figures of merit for the characteristic relative temperature uncertainties obtained from the estimators within that range. 

\begin{figure}
\centering
\includegraphics[width=1.0\columnwidth]{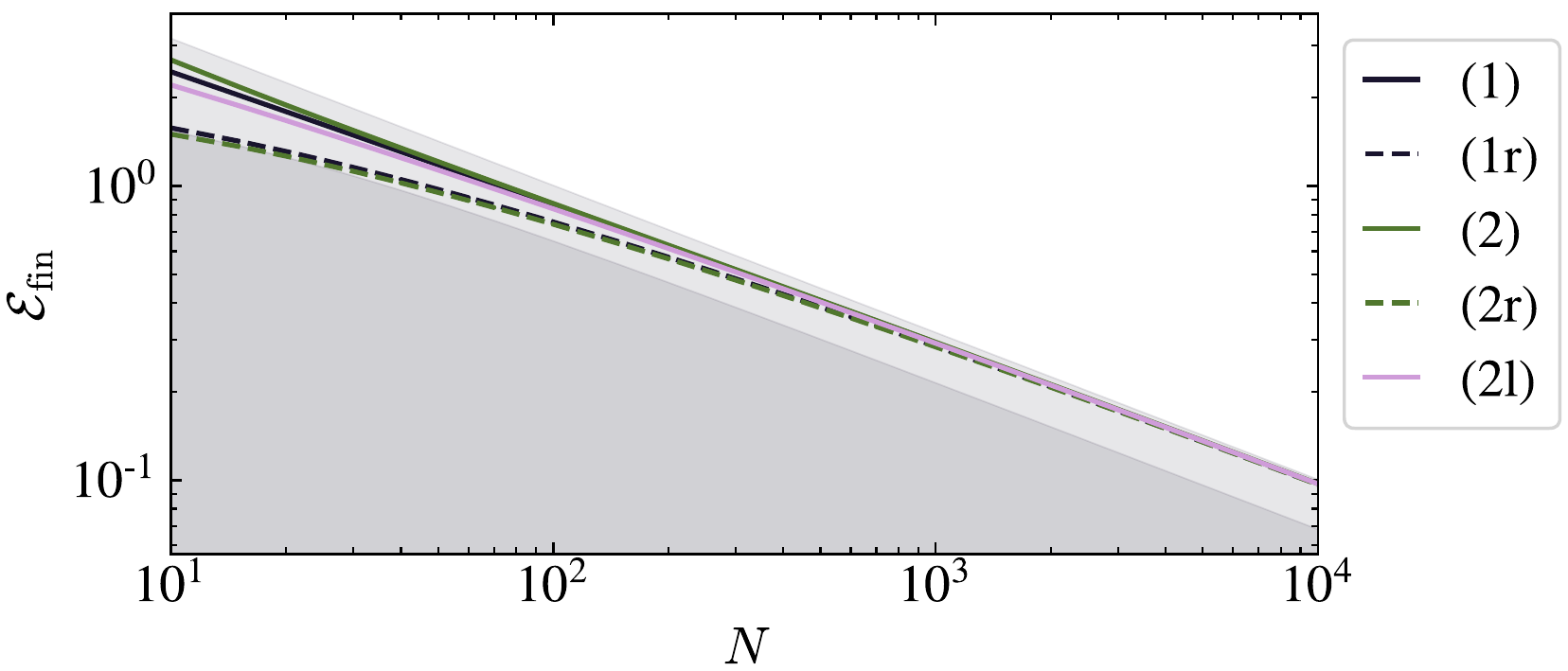}
\caption{\label{fig:scaling_error} (Color online) Outcome-averaged relative errors \eqref{eq:globalE} as a function of probe number $N$, assuming the finite temperature range $k_BT/E\in [0.1,10]$. The listed estimators are compared to the $T$-averaged Cram\'er-Rao bound \eqref{eq:globalCRB} (light shade) and the global error benchmark \eqref{eq:vanTreesBoundEQ} (dark) at thermal equilibrium.}
\end{figure}

In Fig.~\ref{fig:scaling_error}, we plot the $N$-scaling of the overall relative errors \eqref{eq:globalE} averaged over the same range as before, $k_BT/E\in [0.1,10]$. At small $N$, the various estimators show different scaling behaviours, but from about $10^3$ probes onwards, all coincide and scale asymptotically like $1/\sqrt{N}$. This is expected from the Bernstein-von Mises theorem, and it matches the local Cram\'er-Rao bound for the (90\%-equivalent) RMS deviation of unbiased estimators, $\bar\eps_{\text{CRB}}(T) = 3.29/\sqrt{I(T)}$. For explicit comparison, we shall convert it to the relative deviation $\bar\eps_{\text{CRB}} (T)/T$ and take the prior-weighted integral over $[T_1,T_2]$,
\begin{equation}\label{eq:globalCRB}
    \mathcal{E}_{\text{CRB}} = \int_{T_1}^{T_2} \!\! \diff T \frac{P^{(0)}(T) \bar\eps_{\text{CRB}}(T)}{\mathcal{M} T} =  \frac{ 6.58 \, \ln \left( T_2/T_1 \right)}{\pi \sqrt{N} \mathcal{M}}. 
\end{equation}
This asymptotic figure of merit is marked by the light shade in Fig.~\ref{fig:scaling_error}. Indeed, our simulations indicate that the various estimators exhibit decreasing average bias across the chosen $T$-range for sufficiently large $N$, and so the relative errors match \eqref{eq:globalCRB} asymptotically. 

While the Cram\'er-Rao deviation is a good figure of merit in the asymptotic limit, it does not capture the error scaling at finite $N$ where the estimators' biases are significant. Our detailed error analysis as a function of $T$ in Fig.~\ref{fig:comparison equlibrium} has shown that the error measures typically overestimate the actual RMS deviation \eqref{eq:RMSerr} from the true $T$-value. Let us therefore consider the square-root of the prior-weighted integral over the \emph{relative} square deviation $\bar\eps^2_{\text{RMS}}/T^2$ as a global error benchmark, 
\begin{equation}\label{eq:globalErRMS}
    \mathcal{E}_{\text{rRMS}} (\vvtheta) = 3.29 \sqrt{ \cC^{\rm{(2r)}} (\vvtheta) } = \sqrt{ \int_0^{\infty} \!\! \diff T\, P^{(0)}(T) \frac{\bar\eps^2_{\text{RMS}}(T)}{T^2} }.
\end{equation}
Its dimensionless value depends on the chosen estimator through \eqref{eq:RMSerr}, but we can arrive at an estimator-independent lower bound with help of a modified van Trees inequality \cite{Gill1995}, 
\begin{eqnarray} \label{eq:vanTreesBound}
    \mathcal{E}_{\text{rRMS}} &\geq& \frac{3.29}{\sqrt{ \int_0^{\infty}\!\!\diff T\, P^{(0)}(T) T^2 I(T) + I_0 }}, \\
    I_0 &=& \int_0^{\infty}\!\!\diff T\, P^{(0)}(T) T^2 \left[ \partial_T \ln P^{(0)}(T) \right]^2, \nonumber
\end{eqnarray}
see App.~\ref{app:RMSbound}. Here we integrate over the full temperature range for a truly global reference bound, which does not depend on the chosen energy scale $E$ either. The square root in the denominator comprises the Fisher information of the likelihood, integrated over a modified prior distribution, and the offset contribution $I_0$ measuring the $T$-sensitivity of the prior. The latter is rendered negligible by sufficient data, as the former term grows with $I(T) \propto N$. Hence we obtain a temperature-averaged global version of the Cram\'er-Rao bound in the asymptotic limit $N\to\infty$.

However, the global RMS error \eqref{eq:vanTreesBound} is no longer a tight bound, as exemplified by the dark-shaded region in Fig.~\ref{fig:scaling_error}. The errors associated to the relative estimators are well approximated for small $N$, but then consistently underestimated as $N \to \infty$. For this equilibrium case, the integrals in \eqref{eq:vanTreesBound} have explicit analytic solutions, and the global benchmark simplifies to
\begin{equation} \label{eq:vanTreesBoundEQ}
    \mathcal{E}_{\text{rRMS}} \geq \frac{3.29}{\sqrt{ (N+1)\pi^2/8 - N + 1 }} \approx \frac{6.81}{\sqrt{N+9.56}},
\end{equation}
regardless of the qubit energy $E$.

\subsection{Influence of the prior}\label{sec:priorCompare}

\begin{figure}
\centering
\includegraphics[width=1.0\columnwidth]{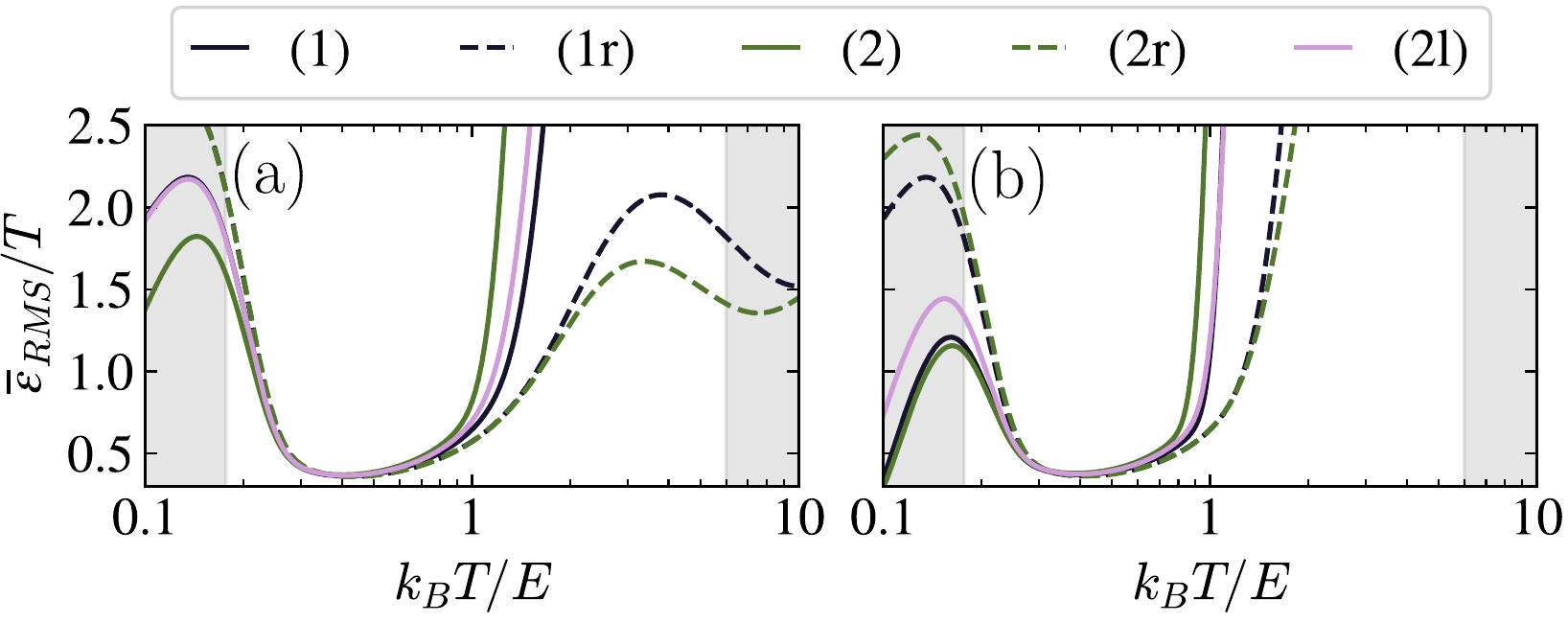}
\caption{\label{fig:comparison prior} (Color online) Influence of the prior distribution on the estimators' accuracies for 200 fully thermalized qubits. We compare the RMS deviations from the true temperature $T$ for (a) the prior $P^{(0)}(T) \propto 1/T$ and (b) the constant prior over the range $k_B T/E \in [0.01,200]$. See Fig.~\ref{fig:comparison equlibrium}(b) for Jeffreys' prior. The grey bars delimit the detectable temperatures.}
\end{figure}

Up to this point in our assessment, we have based our analysis on Jeffreys' prior. However, simpler priors over a finite temperature range are often used in the literature. Here we consider the scale-invariant $1/T$-prior \cite{rubio2020global} and the constant prior \cite{alves2020bayesian} for comparison. The resulting RMS deviations of our estimators are shown in Fig~\ref{fig:comparison prior}(a) and (b), respectively, and should be compared to Fig.~\ref{fig:comparison equlibrium}(b) for Jeffreys' prior. 

The results for the different priors agree only within a small range of temperatures $k_BT/E \lesssim 1$ at which the thermometer scheme is most sensitive. There the posterior distributions after 200 qubits have properly converged to a sharp quasi-Gaussian distribution and ``forgotten'' about the prior, which also explains why estimators all agree. However, the situation changes drastically outside the small optimal $T$-regime: both the $1/T$-prior in Fig~\ref{fig:comparison prior}(a) and the constant prior in (b) amplify the errors. For this reason, Jeffreys' prior is clearly a better choice leading to more accurate results in uninformed qubit thermometry.

\subsection{Nonequilibrium results} \label{sec:noneqResults}

So far, we have studied qubit thermometry at thermal equilibrium and found that the relative mean and median estimators consistently achieve the most accurate estimates and the most reliable errors. This remains the case when considering nonequilibrium probes due to short coupling times $\tau$, which mainly affects the high-temperature behavior of the likelihood function via \eqref{eq:qtau}.

\begin{figure}
\centering
\includegraphics[width=1.0\columnwidth]{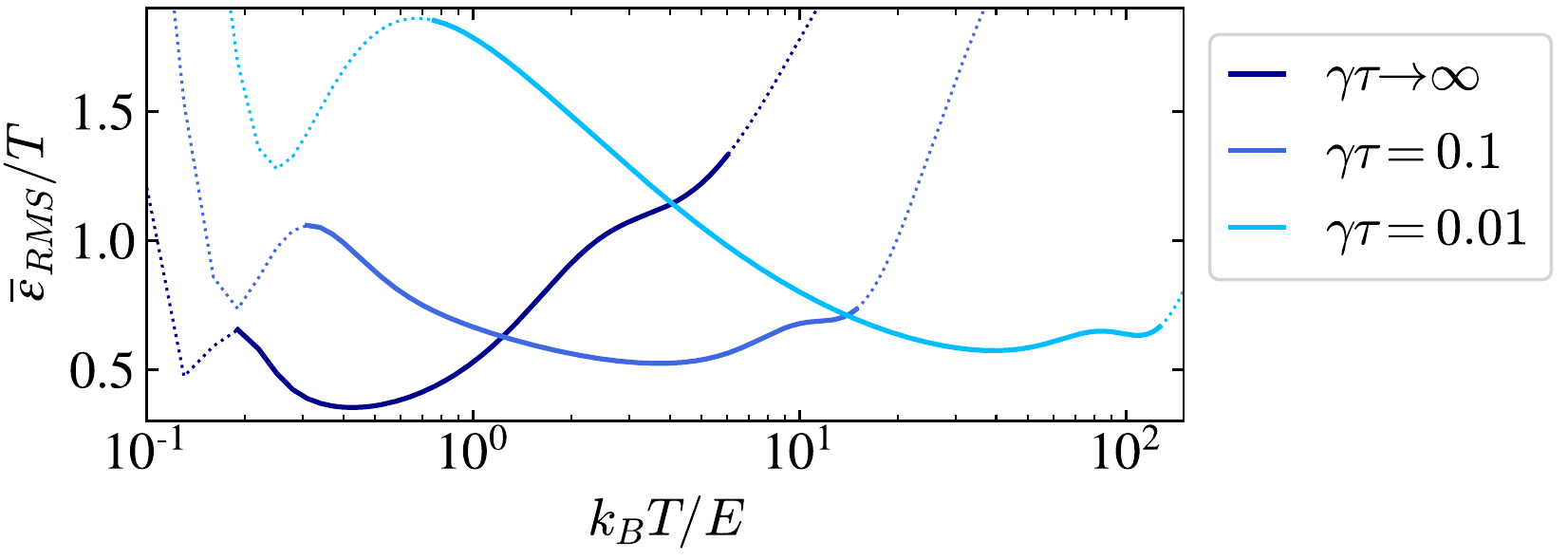}
\caption{\label{fig:comparison non-eqm}RMS deviation relative to temperature for the relative median estimator with 200 qubits. We compare the case of fully thermalized probes (dark blue line) to finite thermalization times, $\gamma \tau = 0.1$ (medium blue) and $\gamma \tau = 0.01$ (light blue). The solid stretches extend over the respective detectable temperature ranges given by \eqref{eq:Tbounds}. }
\end{figure}

Figure \ref{fig:comparison non-eqm} compares our exemplary 200-qubit equilibrium measurement ($\gamma\tau \to \infty$) with short-time non-equilibrium measurements ($\gamma\tau = 0.1$ and $0.01$) in terms of the average RMS deviation for the relative median estimator $\vvtheta^{\rm{(1r)}}$. The solid stretch of each curve extends over the respective range of detectable temperatures from \eqref{eq:Tbounds}. As predicted in Sec.~\ref{sec:sensitivity range}, we find that with shorter $\tau$, the detectable range widens and shifts towards higher temperatures, while the lowest achievable error grows. This trade-off could be controlled and exploited in adaptive estimation protocols.

\section{Conclusion} \label{sec:conclusion}

We have presented an uninformed Bayesian approach to quantum thermometry when a single $N$-probe measurement is performed without prior information about the expected temperature range. Focusing on the case of fully or partially thermalized qubit probes of fixed energy gap, we have compared known and novel temperature estimators and their associated error measures. 

Within the range of temperatures accessible to the measurement, which can be gauged by evaluating the relative entropies of the likelihood function, we have observed striking differences in the attainable accuracy: the widely used mean estimator with its variance-based error measure gives profusely inaccurate predictions and errors at high temperatures. While the median and the recently proposed logarithmic mean estimator improve the accuracy, they do not rectify the exaggerated error.

Our study shows one can achieve a consistently better accuracy and matching error measure by means of either the relative mean or the relative median estimator. The former takes the posterior mean of $1/T$ divided by that of $1/T^2$ as the temperature estimate for a given outcome, while the latter uses the median of the posterior distribution divided by $T$ (and normalized accordingly). Our Bayesian procedure is based on the uninformed Jeffreys prior, which exhibits diverging moments in $T$. Hence the conventional mean and median estimators depend on a high-temperature cutoff, whereas the logarithmic and our relative estimates are always finite. Commonly used alternative priors such as the flat or the scale-invariant one aggravate the divergence problem and generally worsen the estimates.

The temperature-averaged relative errors for most estimators adhere to the likewise averaged Cram\'{e}r-Rao bound in the asymptotic limit $N\to\infty$. At small probe numbers, we find that their scaling is better described by a modified van Trees bound for the relative root-mean-square deviation between estimated and true temperature. This bound is not tight, but universal in that it holds for any temperature range and depends only on $N$.

Finally, we have demonstrated the trade-off between local accuracy and temperature coverage in nonequilibrium thermometry of bosonic reservoirs. By shortening the coupling time of each probe and thereby preventing full thermalization, one can vastly broaden the range of detectable temperatures to higher values at the cost of greater overall errors. This could be a useful control feature for adaptive temperature estimation protocols.

\emph{Acknowledgments.---}
S.S. acknowledges funding from the Swiss National Science Foundation (NCCR SwissMAP). J.B. acknowledges support from the House of Young Talents (HYT) of Siegen University. 


\appendix 

\begin{widetext}

\section{Van Trees inequality for the global relative RMS error}\label{app:RMSbound}

Here we derive the estimator-independent lower bound \eqref{eq:vanTreesBound} for the global relative RMS error \eqref{eq:globalErRMS}, similar to the original van Trees inequality for the \emph{absolute} RMS error averaged over the prior. To this end, we adopt the derivation laid out in \cite{Gill1995}. 

For convenience and generality, we shall work with arbitrary (discrete or continuous) vector-valued measurement outcomes $\vx$, introducing a suitable integral measure $\int\diff\vx$ over the data space. The likelihood function $P(\vx|T)$ is then a positive probability distribution on the data space, normalized to $\int \diff \vx \, P(\vx|T) = 1$ for all values of $T$. In our specific case, replace $\vx \to n$ and $\int\diff \vx \to \sum_{n=0}^N$. 

Now consider the integral
\begin{equation} \label{eq:J}
    \mathcal{J} = \int_0^\infty \!\! \diff T \int \!\! \diff \vx \left[ \theta(\vx) - T \right] \frac{\diff}{\diff T}\left[ P(\vx|T) P^{(0)}(T) \right],
\end{equation}
with $\theta(\vx)$ the outcomes of a given estimator $\vtheta$. We admit arbitrary positive $T$-values here; restrictions to a finite interval should be encoded explicitly in the chosen prior $P^{(0)}(T) \geq 0$, normalized as $\int_0^\infty \diff T \, P^{(0)}(T) = 1$. Introducing the scalar product $\la f,g\ra = \int_0^\infty \diff T\int\diff \vx \, f(\vx,T) g(\vx,T)$ for real-valued functions over the combined $(\vx,T)$ space, we can rewrite $\mathcal{J} = \la f,g \ra$ with
\begin{equation}
    f(\vx,T) = \sqrt{P(\vx|T)P^{(0)}(T)} \frac{\theta(\vx)-T}{T}, \qquad g(\vx,T) = T \sqrt{P(\vx|T)P^{(0)}(T)} \frac{\diff}{\diff T} \ln \left[ P(\vx|T)P^{(0)}(T) \right],
\end{equation}
assuming that both functions have finite norm. By virtue of the Cauchy-Schwarz inequality, we then get
\begin{equation}
    \la f,f\ra = \int_0^\infty\!\!\diff T \int\!\!\diff\vx \left[\frac{\theta(\vx)-T}{T} \right]^2 P(\vx|T) P^{(0)}(T) = \cC^{\rm{(2r)}} (\vtheta) \leq \frac{\mathcal{J}^2}{\la g,g\ra}. \label{eq:CSU}
\end{equation}
Notice that the squared norm of $f$ is precisely the average cost associated to the relative mean estimator, as defined in the main text for $\vx \equiv n$. The other norm can be expressed as
\begin{eqnarray}
    \la g,g\ra &=& \int_0^\infty\!\!\diff T \int\!\!\diff\vx \, P(\vx|T) P^{(0)}(T) T^2 \left[ \frac{\diff}{\diff T} \ln P(\vx|T) + \frac{\diff}{\diff T} \ln Q(T) \right]^2 \nonumber \\
    &=& \int_0^\infty\!\!\diff T \, P^{(0)}(T) T^2 \left\{ \int \!\!\diff \vx \, P(\vx|T) \left[ \frac{\diff \ln P(\vx|T)}{\diff T} \right]^2 +  \left[ \frac{\diff \ln P^{(0)}(T)}{\diff T} \right]^2 + 2 \frac{\diff \ln P^{(0)}(T)}{\diff T} \int \!\!\diff \vx \frac{\diff}{\diff T} P(\vx|T) \right\}\nonumber \\
    &=& \int_0^\infty\!\!\diff T \, P^{(0)}(T) T^2 I(T) + I_0 + 0. \label{eq:gNorm}
\end{eqnarray}
In the last step, we identify the Fisher information $I(T)$ of the likelihood and the prior term $I_0$, and we notice that the last term in the second line vanishes as we exchange the $\vx$-integration with the $T$-derivative. What remains is to simplify the expression \eqref{eq:J} for $\mathcal{J}$ by means of partial integration, 
\begin{equation}
    \mathcal{J} = \int \!\! \diff \vx \left\{ \big[ P^{(0)}(T) P(\vx|T) (\theta(\vx)-T) \big]_0^\infty + \int_0^\infty \!\! \diff T \,  P(\vx|T) P^{(0)}(T) \right\} = 1. \label{eq:J2}
\end{equation}
For the boundary terms to vanish, we make the assumption that $P^{(0)}(T) \xrightarrow{T\to 0} 0$ and $ T P^{(0)}(T) \xrightarrow{T\to \infty} 0$, which is indeed the case for Jeffreys' prior in our scenario. The desired bound \eqref{eq:vanTreesBound} in the main text follows by inserting \eqref{eq:J2} and \eqref{eq:gNorm} into \eqref{eq:CSU}, taking the square root, and multiplying by the factor $3.29$ for consistency with our 90\% credibility convention.
For the equilibrium case ($\gamma\tau \to\infty$), it turns out that the unrestricted integrals can be evaluated explicitly,
\begin{equation}
    \int_0^\infty \!\!\diff T \, P^{(0)}(T) T^2 I(T) = N \left( \frac{\pi^2}{8} - 1 \right), \quad
    I_0 =\frac{\pi^2}{8} +1.
\end{equation}

\end{widetext}

\end{document}